\documentstyle[12pt]{article}
\input epsf
\newcommand{\be}{\begin{eqnarray}}
\newcommand{\ee}{\end{eqnarray}}

\newcommand{\la}[1]{\label{#1}}
\newcommand{\eq}[1]{eq.~(\ref{#1})}

\newcommand{\ur}[1]{~(\ref{#1})}

\newcommand{\Eq}[1]{Eq.~(\ref{#1})}

\newcommand{\ra}[1]{(\ref{#1})}
  \def\beq{\begin{equation}}
  \def\eeq{\end{equation}}
  \def\beqr{\begin{eqnarray}}
  \def\eeqr{\end{eqnarray}}
  \def\nn{\nonumber\\}

 \def\Sp{\mbox{Sp}}

 \def\bra#1{{\langle#1\vert}}
 \def\ket#1{{\vert#1\rangle}}
 
 \def\Dirac#1{#1\hskip-6pt/}
 \def\dd{\Dirac\partial}

\begin{document}
\rightline{RUB-TPII-15/96}
%
\vspace{1cm}
\begin{center}
{\bf\Large
Transverse Spin Distribution Function of Nucleon in Chiral Theory
} \\
\vspace{1cm}
{\bf\large
P.V.~Pobylitsa$^1$, M.V.~Polyakov$^2$ }\\
\vspace{0.25cm}
{\it Petersburg Nuclear Physics Institute, Gatchina,
St.Petersburg 188350, Russia} \\[.1cm]
{\it Institut f\"ur Theoretische Physik II,
Ruhr-Universit\"at Bochum, D-44780 Bochum, Germany}
\end{center}
\vspace{1.5cm}
\begin{abstract}
\noindent
At large $N_c$ the nucleon can be viewed as a soliton of the effective
chiral lagrangian. This picture of nucleons allows
a consistent nonperturbative calculation of the leading-twist parton
distributions at a low normalization point. We derive general formulae for
the transverse spin quark distribution $h_1(x)$ in
the chiral quark-soliton model.
We present numerical estimates and compare them to the results
obtained in other models.
\end{abstract}
%
%
\vfill
\rule{5cm}{.15mm}
\\
\noindent
{\footnotesize $^{\rm 1}$ E-mail: pavelp@hadron.tp2.ruhr-uni-bochum.de}
\\ {\footnotesize $^{\rm 2}$ E-mail: maxpol@thd.pnpi.spb.ru} \\

\newpage

{\bf 1.}
The transverse polarization distribution of quarks
introduced by Ralston and Soper \cite{RS} and usually called
$h_1(x)$ has gained an increasing amount of attention of theorists
\cite{AM,JJ,JaffeJi1,JaffeJi2,Ji,CPR}.
The perspective of the experimental measurement of $h_1(x)$
\cite{RHIC,COMPASS}
has stimulated attempts to calculate $h_1(x)$ in various models of strong
interactions \cite{JaffeJi1,JaffeJi2,Ioffe,Barone}.

Recently a new approach to the calculation of quark
distribution functions has been developed \cite{DPPPW} within
the context of the effective chiral quark-soliton model of nucleon
\cite{DPP}.
In this paper we apply this approach to the
calculation of  the transverse spin distribution function $h_1(x)$.
Although in reality the number of colours $N_c=3$, the academic limit
of large $N_c$ is known to be a useful guideline. At large $N_c$ the
nucleon is heavy and can be viewed as a classical soliton of the
pion field \cite{Witten,ANW}. In this paper we work with the
effective chiral action given by the
functional integral over quarks in the background pion field
\cite{DE,DSW,DP}:
\be
\exp\left(iS_{\rm eff}[\pi(x)]\right) &=&
\int D\psi D\bar\psi \; \exp\left(i\int d^4 x\,
\bar\psi(i\dd - MU^{\gamma_5})\psi\right), \nonumber
\ee
\beq
U \; = \; \exp\left[i\pi^a(x)\tau^a\right], \hspace{1cm}
U^{\gamma_5} \; = \; \exp\left[i\pi^a(x)\tau^a\gamma_5\right] \; =
\; \frac{1+\gamma_5}2 U
+ \frac{1-\gamma_5}2 U^\dagger.
\la{FI}
\eeq
Here $\psi$ is the quark field, $M$ is the effective quark mass
which is due to the spontaneous breakdown of chiral symmetry
and $U$ is the $SU(2)$ chiral pion
field. The effective chiral action given by \eq{FI} is known to
contain automatically the Wess--Zumino term and the four-derivative
Gasser--Leutwyler terms, with correct coefficients.
\Eq{FI} has been derived from the instanton model of the QCD vacuum
\cite{DP,DP1}, which provides a natural mechanism of chiral
symmetry breaking and enables one to express the dynamical mass $M$ and the
ultraviolet cutoff intrinsic in \eq{FI} through the
$\Lambda_{QCD}$ parameter. It should be mentioned that \eq{FI} is
of a general nature: one need not believe in instantons
and still can use \eq{FI}.

An immediate application of the effective chiral theory \ur{FI} is the
quark-soliton model of baryons of ref.~\cite{DPP}, which is in the
spirit of the earlier works~\cite{KaRiSo,BiBa}.
According to this model nucleons
can be viewed as $N_c$ (=3) ``valence" quarks bound by a self-consistent
pion field (the ``soliton") whose energy coincides in fact
with the aggregate energy of the quarks of the negative-energy Dirac
continuum. Similarly to the Skyrme model large $N_c$ is needed as a
parameter to justify the use of the mean-field approximation,
however, the $1/N_c$ corrections can be and in some cases have been computed
\cite{WY,Chr,Chr2,Kim,Kim1}.  The quark-soliton model of nucleons
developed in ref.~\cite{DPP} includes a collective-quantization
procedure to deal with the rotational excitations of the quark-pion
soliton.

Turning to the calculation of
the transverse spin distribution function $h_1(x)$
we note
that the model possesses all features needed for a successful description
of the nucleon parton structure: it is an essentially quantum
field-theoretical relativistic model with explicit quark degrees of
freedom, which allows an unambiguous identification of quark as well as
antiquark distributions in the nucleon. This should be contrasted to the
Skyrme model where it is not too clear how to define quark and antiquark
distributions.

%
\vspace{0.5cm}

{\bf 2.}
The transverse spin quark distribution function for flavour $f$
is defined as follows \cite{JaffeJi1, JaffeJi2}
\begin{equation}
h_1^f(x) =  \int \frac{d\lambda}{4\pi} e^{i\lambda x}
\bra{P,S_T} \bar\psi_f(0) {\Dirac n} \gamma_5
{\Dirac S}_T \psi_f(\lambda n)
\ket{P,S_T} \,.
\label{h-T-definition}
\end{equation}
Below we work in the nucleon rest frame
where the transverse nucleon spin $S_T$ and the light-cone vector $n$
can be chosen as follows:
\begin{equation}
n^\mu = \frac{1}{M_N} (1,-1,0,0)\,,\qquad
S_T^\mu = (0,0,0,1) \, .
\end{equation}

The operators in eq.~(\ref{h-T-definition})
depend on the QCD
normalization point $\mu$. In contrast to QCD,
the effective chiral field theory is nonrenormalizable and
contains an explicit ultraviolet cutoff.
In the instanton vacuum model this cutoff
appears as the inverse average instanton size:
$\bar\rho^{-1}\approx 600$~MeV \cite{DP1,DPW}. The results
obtained below thus refer to a low point normalization point $\mu$ of the
order of $600$~MeV.

Now we can rewrite (\ref{h-T-definition}) in the form
\beqr
&&
h_1^f(x)
= -  \frac{1}{4\pi}
\int\limits_{-\infty}^\infty dz^0  e^{ixM_Nz^0}
\nn
&&
\times
\bra{P,S_T} \psi_f^+(0) (1+\gamma^0\gamma^1)\gamma_5
\gamma^3
\psi_f(z) \ket{P,S_T}
\Bigr|_{z^1=-z^0,\>z^2=z^3=0} \, .
\label{h1-definition}
\eeqr

Let us remind the reader how the nucleon is described in the
effective low-energy theory \ra{FI}.
Integrating out the quarks in \ra{FI} one
finds the effective chiral action,
\be
S_{\rm eff}[\pi^a(x)] &=& -N_c\,\Sp\log D(U)\,, \hspace{1cm}
D(U) \;\;= \;\; i\partial_0 - H(U),
\label{SeffU}
\ee
where $H(U)$ is the one-particle Dirac hamiltonian,
\be
H(U) &=& - i\gamma^0\gamma^k \partial_k + M\gamma^0 U^{\gamma_5} \,,
\la{hU}
\ee
and $\mbox{Sp}\ldots$ denotes a functional trace.

For a given time-independent pion field $U=\exp(i\pi^a({\bf x})\tau^a)$
one can determine the spectrum of the Dirac hamiltonian,
\be
H\Phi_n &=& E_n \Phi_n.
\la{Dirac-equation}
\ee
It contains the upper and lower Dirac continua (distorted by the
presence of the external pion field), and, in principle, also
discrete bound-state level(s), if the pion field is strong enough. If
the pion field has unity winding number, there is exactly one
bound-state level which travels all the way from the upper to the lower
Dirac continuum as one increases the spatial size of the pion field
from zero to infinity \cite{DPP}. We denote the energy of the discrete
level as $E_{\rm lev},\;\;-M\leq E_{\rm lev}\leq M$.  One has to occupy
this level to get a non-zero baryon number state. Since the pion field is
colour blind, one can put $N_c$ quarks on that level in the antisymmetric
state in colour.

The limit of large $N_c$ allows us to use the mean-field approximation
to find the nucleon mass.
To get the nucleon mass one has to add
$N_cE_{\rm lev}$ and the energy of the pion field. Since the effective
chiral lagrangian is given by the determinant \ur{SeffU} the energy of the
pion field coincides exactly with the aggregate energy of the lower Dirac
continuum, the free continuum subtracted. The self-consistent pion
field is thus found from the minimization of the functional \cite{DPP}

\be
M_N &=& \min_U \; N_c\left\{E_{\rm lev}[U]
\; + \; \sum_{E_n<0}(E_n[U]-E_n^{(0)})\right\}.
\la{nm}
\ee
From symmetry considerations one looks for the minimum in a hedgehog
ansatz:
\beq
U_c({\bf x}) \; = \; \exp\left[i\pi^a({\bf x})\tau^a\right]
\; = \; \exp\left[i n^a \tau^a P(r)\right],
\hspace{1cm} r \; = \; |{\bf x}|,
\hspace{1cm} {\bf n} \; = \; \frac{{\bf x}}{r} ,
\la{hedge}
\eeq
where $P(r)$ is called the profile of the soliton.

The minimum of the energy \ur{nm} is degenerate with respect to
translations of the soliton in space and to rotations of the soliton
field in ordinary and isospin space. For the hedgehog field \ur{hedge} the
two rotations are equivalent. Quantizing slow rotations of the saddle-point
pion field \cite{ANW,DPP} leads to
the projection on a nucleon with given spin ($S_3$) and isospin
($T_3$) components
which includes the integration over all spin-isospin
$SU(2)$ orientation matrices $R$ of the soliton,
\be
\langle S=T,S_3,T_3|\ldots| S=T,S_3,T_3\rangle
&=& \int dR\;\phi^{\ast\;S=T}_{S_3T_3}(R) \; \ldots \;
\phi^{S=T}_{S_3T_3}(R)\,.
\la{spisosp}
\ee
Here $\phi^{S=T}_{S_3T_3}(R)$ is the rotational wave function of the
nucleon given by the Wigner finite-rotation matrix \cite{ANW,DPP}:

\be
\phi^{S=T}_{S_3T_3}(R) &=&
\sqrt{2S+1}(-1)^{T+T_3}D^{S=T}_{-T_3,S_3}(R).
\la{Wigner}
\ee
\vspace{0.5cm}

{\bf 3.}
 \Eq{h-T-definition} contains
a matrix element of a non-local quark bilinear operator in
the nucleon state with definite 4-momentum $P$ and spin and isospin
components. According to \cite{DPPPW}
one can write a
general expression for such matrix elements; the time dependence of the
quark operators is accounted for by the energy exponents.
Taking nucleon at rest, we can write this matrix element
as a sum over all occupied quark states
\be
\lefteqn{
\bra{S_3,T_3}\, \psi^\dagger_{fi}(x^0,{\bf x}) \,
\psi^{gj}(y^0,{\bf y}) \,
\ket{S_3,T_3} \;\;\; = \;\;\; 2M_NN_c \int d^3{\bf X}
\int dR \; \phi_{S_3T_3}^\ast(R)}
\nn
&&
\times
\sum\limits_{\scriptstyle n\atop \scriptstyle{\rm
occup.}}\exp[iE_n(x^0-y^0)] \; \Phi_{n,f^\prime i}^\dagger({\bf
x-X}) \, (R^\dagger )^{f^\prime}_f R_{g^\prime}^g
\, \Phi_n^{g^\prime j}({\bf y-X}) \, \phi_{S_3T_3}(R)\,.
\la{annihilate}
\ee
Here we have written
explicitly all the flavour ($f,g=1,2$) and the Dirac ($i,j=1,...,4$)
indices for clarity.
The functions $\Phi_n$ are eigenstates of energy $E_n$ of the Dirac
hamiltonian \ur{hU} in the external (self-consistent) pion field $U_c$.
Summation over colour indices is implied in the quark bilinears, hence the
factor $N_c$.

Applying the general formula of the effective chiral
theory (\ref{annihilate}) to the calculation of the
transverse spin distribution function (\ref{h1-definition})
we obtain in the leading order of the $1/N_c$ expansion
\beqr
&&
h_1^f(x)
= - \frac{N_c M_N}{2\pi} \int dR |\phi_{T^3 S^3}(R)|^2
\int\limits_{-\infty}^\infty dz^0  e^{ixM_Nz^0}
\sum\limits_{\scriptstyle n\atop \scriptstyle
{\rm occup.}}
e^{-iE_n z^0}
\nn
&&
\times
\bra n R^\dagger T^f R
(1+\gamma^0\gamma^1)\gamma_5 \gamma^3
\exp(-iz^0 p^1)
\ket n  \, .
\la{h-T-f-general-leading-Nc}
\eeqr
Here $T^f$ is the matrix projecting onto the flavour $f$.
We use notation $p^k=-i\partial_k$ for the operator of the 3-momentum.
One can easily check that in the leading order of the $1/N_c$
expansion only the {\em isovector} polarized distribution
survives:
\beqr
&&
h_1^u(x) - h_1^d(x)
= -  \frac{N_c M_N}{2\pi} \int dR |\phi_{T^3 S^3}(R)|^2
\int\limits_{-\infty}^\infty dz^0  e^{ixM_Nz^0}
\sum\limits_{\scriptstyle n\atop \scriptstyle
{\rm occup.}}
e^{-iE_n z^0}
\nn
&&
\times
\bra n R^\dagger \tau^3 R
(1+\gamma^0\gamma^1)\gamma_5 \gamma^3
\exp(-iz^0 p^1)
\ket n   \, .
\la{h-T-nonsinglet-1}
\eeqr
The integral over the orientation matrix $R$ can be
easily computed. As a result we obtain
\beqr
&&
h_1^u(x) - h_1^d(x)
=  \frac{N_c M_N}{6\pi}
\int\limits_{-\infty}^\infty dz^0  e^{ixM_Nz^0}
\sum\limits_{\scriptstyle n\atop \scriptstyle
{\rm occup.}}
e^{-iE_n z^0}
\nn
&&
\times
\bra n
\tau^3
(1+\gamma^0\gamma^1)\gamma_5 \gamma^3
\exp(-iz^0 p^1)
\ket n  \, .
\la{h-T-nonsinglet-2}
\eeqr
Taking into account the hedgehog symmetry \ra{hedge}
we can replace here
$1\leftrightarrow 3$. Passing to the momentum representation
and integrating over $z^0$, we obtain
the expression convenient for practical calculations:
\beqr
&&
h_1^u(x) - h_1^d(x)
=    \frac{N_c M_N}{6\pi}
\int \frac{d^2p_\perp }{(2\pi)^2}
\sum\limits_{\scriptstyle n\atop \scriptstyle
{\rm occup.}}
\nn
&&
\times
\bra n {\bf p} \rangle
(1+\gamma^0\gamma^3)\gamma_5\gamma^1\tau^1 \langle {\bf p} \ket n
\Bigr|_{p^3=xM_N-E_n} \, .
\la{h-T:nonsinglet:general-2}
\eeqr

It is well known that the transverse spin distribution function $h_1$
satisfies the sum rule \cite{JaffeJi1, JaffeJi2}
\begin{equation}
\int\limits_{-1}^{1} dx\, h_1^f(x)= g_T^f
\la{T-sum-rule}
\end{equation}
where $g_T^f$ is the tensor charge of the nucleon defined as
follows
\begin{equation}
\bra{P,S} {\bar\psi}_f [\gamma_\mu,\gamma_\nu] \psi_f \ket{P,S}
= g_T^f  \bar u(P,S) [\gamma_\mu,\gamma_\nu] u(P,S)\, .
\la{g-T-definition}
\end{equation}
Here $u(P,S)$ is the standard Dirac spinor describing the nucleon
with momentum $P$ and spin $S$.

Substituting expression for $h_1(x)$ \ra{h-T:nonsinglet:general-2}
into the lhs of the sum rule \ra{T-sum-rule} and
extending the integral over $x$ to the whole real axis
(which is justified by the large $N_c$ limit \cite{DPPPW})
we obtain
\begin{equation}
\int\limits_{-1}^{1} dx\, h_1^f(x)
=  \frac{N_c}{3}
\sum\limits_{\scriptstyle n\atop \scriptstyle
{\rm occup.}}
\bra n \gamma_5 \gamma^1 \tau^1 \ket n  \, .
\end{equation}
The result in the rhs coincides with the
expression obtained in paper \cite{Kim} for the tensor charge of
nucleon so that the sum rule \ra{T-sum-rule} automatically holds
in the effective chiral theory.

\vspace{0.5cm}

{\bf 4.}
In eq.~\ra{h-T:nonsinglet:general-2}
one has to sum over all occupied levels of the Dirac Hamiltonian,
including both negative Dirac continuum and discrete level.

In contrast to the longitudinal spin quark distribution function
$g_1(x)$ whose Dirac continuum contribution is ultraviolet
divergent \cite{DPPPW}, in the transverse case
the continuum contribution is ultraviolet finite. This can be seen
from the gradient expansion for $h_1(x)$ where the first nonvanishing
term appears in higher orders compared to the gradient expansion
of $g_1(x)$. This also hints that the continuum contribution
to $h_1(x)$ is relatively small,
whereas the continuum contribution to
the longitudinal spin distribution function $g_1(x)$
is rather sizable \cite{DPPPW}.
An additional argument in favour of the suppression
of the continuum contribution to $h_1$
comes from
the numerical calculation of the tensor charge performed in
\cite{Kim,Kim1},
which shows that the contribution of the Dirac continuum
is essentially smaller than the contribution of the discrete level.
Since the sum rule \ra{T-sum-rule} holds in our model, it is natural to
expect that the discrete level gives the dominant contribution
also to the quark distribution function $h_1(x)$.

The bound-state level occurs \cite{DPP} in the grand spin $K=0$ and parity
$\Pi=+$ sector of the Dirac hamiltonian \ra{hU}.
In that sector the eigenvalue equation takes the form

\be
\left(\begin{array}{cc}
 M \cos P(r) &
{\displaystyle -\frac{\partial}{\partial r}-\frac{2}{r} - M \sin P(r)}\\
{\displaystyle \frac{\partial}{\partial r} - M \sin P(r)} & - M \cos P(r)
\end{array}\right)
\left(\begin{array}{c}
h_0(r) \rule[-.5em]{0cm}{2em} \\ j_1(r) \rule[-.5em]{0cm}{2em}
\end{array}\right)
&=& E_{\rm lev}
\left(\begin{array}{c}
h_0(r) \rule[-.5em]{0cm}{2em} \\ j_1(r) \rule[-.5em]{0cm}{2em}
\end{array}\right).
\la{H-K-0:Pi-plus}
\ee
We assume that the radial wave functions are normalized by the
condition

\be
\int\limits_{0}^{\infty} dr \, r^2  \, [h_0^2(r) + j_1^2(r)] &=& 1.
\ee
Introducing the radial wave functions in the momentum representation
\be
h(k) &=& \int\limits_{0}^{\infty} dr \, r^2\, h_0(r)
\sqrt{ \frac{k}{r}} J_{\frac12}(kr)\,,
\hspace{1.2cm}
j(k) \;\; = \;\; \int\limits_{0}^{\infty} dr \, r^2\, j_1(r)
\sqrt{ \frac{k}{r}} J_{\frac32}(kr)\,,
\ee
we can write the expression for the level contribution to $h_1$
in the form
\be
\left[ \, h_1^u(x) \, - \, h_1^d(x) \, \right]_{\rm lev}
&=& \frac13  N_c M_N
\!\!\int\limits_{|xM_N - E_{\rm lev}|}^{\infty}\!\!
\frac{dk}{2k} \; \Biggl\{ h^2(k)
+ \frac{(xM_N - E_{\rm lev})^2}{k^2}  j^2(k)
\nn
&&
- 2 \frac{(xM_N - E_{\rm lev})}{k} h(k) j(k)\Biggr\}.
\la{valence2}
\ee

\vspace{0.5cm}

{\bf 5.}
Recently the following inequality has been suggested
by Soffer  \cite{Soffer}
\begin{equation}
q^f(x) + g_1^f(x) \ge 2 |h_1^f(x)| \, .
\la{Soffer-inequality}
\end{equation}
Its status in QCD has been studied in \cite{GJJ}.
Let us check this inequality in the chiral quark soliton model.
A delicate point is that in the leading order of the $1/N_c$
expansion the unpolarized quark distribution
$q^f$ is saturated by the singlet part whereas
$g_1^f$ and $h_1^f$ get the main contribution from their nonsinglet parts.
Therefore in the large $N_c$ approach the above inequality can be
rewritten as follows:
\begin{equation}
[q^u(x) + q^d(x)] - |g_1^u(x) - g_1^d(x)|  \ge
2|h_1^u(x) - h_1^d(x)| \, .
\la{inequality-3}
\end{equation}
Note that the structure of the expressions for all quark distribution
function in the effective quark soliton model
entering inequality~\ra{inequality-3} is essentially the same.
They contain a sum of diagonal matrix elements over
occupied states of the Dirac hamiltonian, differing
only by the spin and isospin matrices $\Gamma$ appearing in these
matrix elements:
\begin{eqnarray}
& [q^u(x) + q^d(x)]\,, & \Gamma_q = (1+\gamma^0\gamma^3)\, ,
\nonumber\\
& [g_1^u(x) - g_1^d(x)]\,, & \Gamma_L = - \frac{1}{3}
(1+\gamma^0\gamma^3)\gamma_5 \tau^3  \, ,
\nonumber\\
& [h_1^u(x) - h_1^d(x)]\,, & \Gamma_T =  \frac{1}{3}
(1+\gamma^0\gamma^3)\gamma_5 \gamma^1 \tau^1
\end{eqnarray}
(the explicit expressions for $q^f(x)$ and $g_1(x)$
can be found in ref.~\cite{DPPPW}).
Taking into account the following inequalities
(understood in the matrix sense)
\begin{equation}
|(1+\gamma^0\gamma^3)\gamma_5 \tau^3| \le 1+\gamma^0\gamma^3 \, ,
\end{equation}
\begin{equation}
|(1+\gamma^0\gamma^3)\gamma_5 \gamma^1\tau^1| \le 1+\gamma^0\gamma^3
\end{equation}
one can easily see that the Soffer inequality \ra{Soffer-inequality}
holds for the contribution of each separate occupied level.

One should keep in mind that in our model
the quark distribution functions
$q(x)$ and $g_1(x)$ are ultraviolet divergent and in principle
the regularization can violate the above argument.
However, at least in the limit of large cutoff we can
prove the validity of the Soffer inequality.
Indeed, function $h_1(x)$ is ultraviolet finite
whereas $q(x)$ and $g_1(x)$ are ultraviolet divergent.
The analysis of these divergences has been performed in
ref.~\cite{DPPPW} and it follows from this analysis that
the ultraviolet divergence of
the  combination $[q^u(x) + q^d(x)]-|g_1^u(x) - g_1^d(x)|$
is given by an explicitly positive expression so that
in the limit of large cutoff this expression definitely becomes larger
than the ultraviolet finite $h_1(x)$, which proves that the Soffer
inequality holds in this limit.

\vspace{0.5cm}

{\bf 6.}
We have calculated numerically the level contribution to the
transverse isovector quark distribution. In this calculation we use
the parameters obtained by using
a variational estimate of the profile of the pion-field soliton
(see \eq{hedge}) performed in ref. \cite{DPP} yielding
for $M=350$~MeV
\beq
P(r) \;\; = \;\; -2\;\arctan\left(\frac{r_0^2}{r^2}\right) ,
\hspace{1cm}
r_0 \;\; \approx \;\; 1.0/M ,
\hspace{1cm} M_N \;\; \approx \;\; 1170\;{\rm MeV}.
\la{varprof}\eeq
This profile function has a correct behaviour at small and large distances
and is stable with respect to small perturbations. In our numerics we
use the analytic profile~\ra{varprof}.

The results of our calculations for $g_1$ and $h_1$
are presented at Fig.1.
The antiquark distributions are shown at Fig.2.
From the latter plot we can see that the antiquark transverse spin
distribution is essentially smaller than the corresponding longitudinal
distribution which is related to the fact that the longitudinal
distribution gets a large contribution from the quarks of the Dirac
continuum \cite{DPPPW} in contrast to the transverse distribution.

Comparing our results to the calculations in other models --
bag model \cite{JaffeJi1,JaffeJi2}, QCD sum rules
\cite{Ioffe}, chiral chromodielectric model
\cite{Barone}, one can see that the main source of difference is the
$1/N_c$ approximation used in our calculations. As it was mentioned
above, in the leading order of the $1/N_c$ expansion we have
$h_1^u(x)=- h_1^d(x)$. This relation is much more general
than the model we used and also holds in the large $N_c$ QCD.
It is well known that typically the large $N_c$ counting
is a good guide line for understanding nonperturbative physics.
In order to understand the reliability of the $1/N_c$
expansion in the case of $h_1$ it were instructive to compute
the next-order $1/N_c$ corrections. Unfortunately,
the calculation of these corrections in our model is rather complicated
in the case of the quark distribution functions.
However, these $1/N_c$ corrections have been computed
for the tensor charge \cite{Kim,Kim1} and this calculation shows that
these corrections really enhance $g_T^u$ and suppress $g_T^d$.  This
means the $1/N_c$ corrections of our model can reconcile the large
$N_c$ approach with the model calculations of
\cite{JaffeJi1,JaffeJi2,Ioffe,Barone}.

\vspace{0.5cm}
{\large\bf Acknowledgements} \\[.3cm]
This work has been supported in part by Alexander von Humboldt 
Foundation, by the NATO Scientific Exchange grant OIUR.LG 951035
and by the RFBR grant 95-02-03662.  We acknowledge the hospitality of 
Bochum University.  We appreciate the multiple discussions with 
D.~Diakonov, H.-C.~Kim, V.~Petrov and C.~Weiss.  It is a pleasure to 
thank K. Goeke for encouragement and multiple help.

\newpage
\begin{figure}
 \vspace{-1cm}
\epsfxsize=16cm
\epsfysize=15cm
\centerline{\epsffile{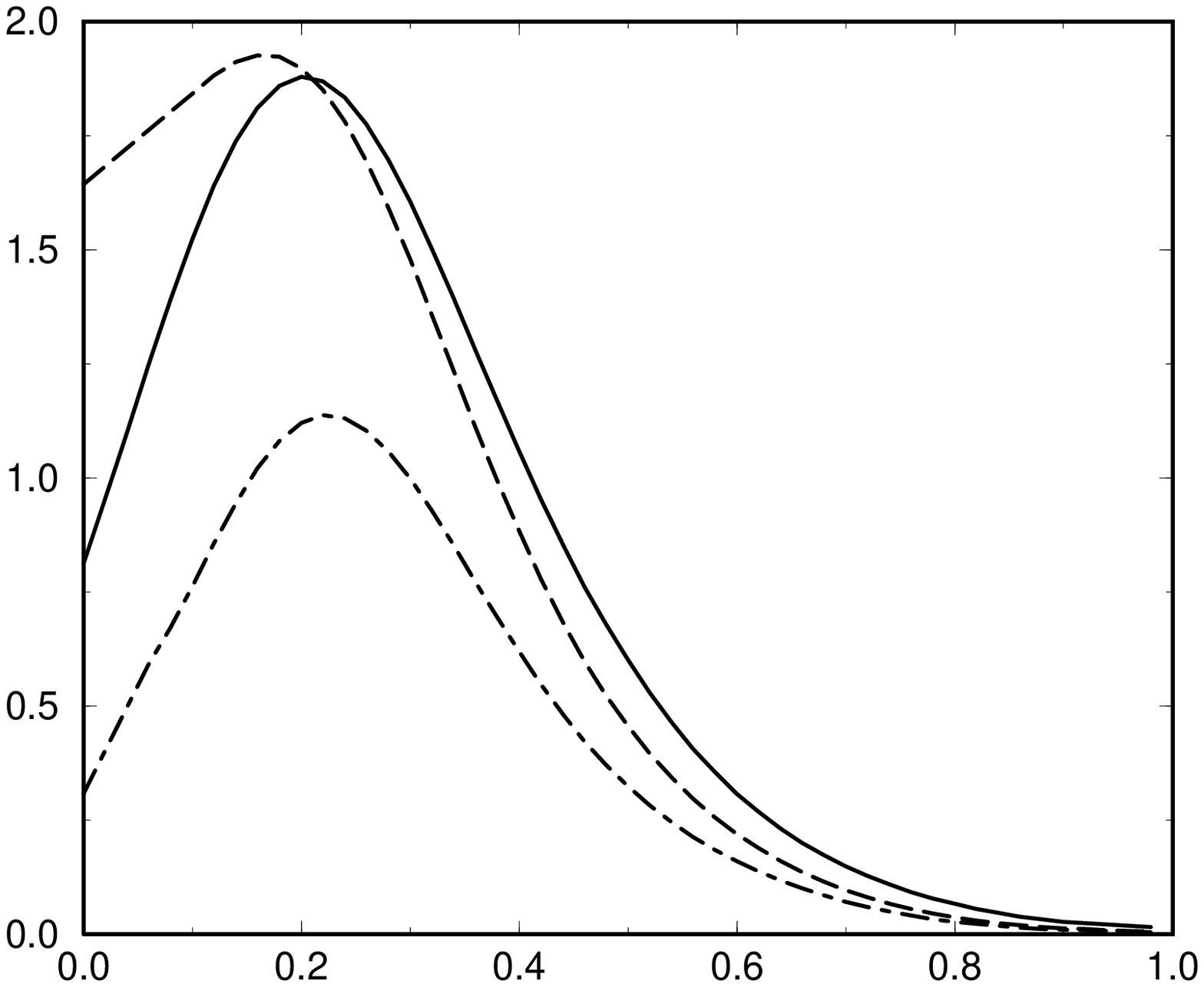}}
\caption[]{
The transverse spin quark distribution function
$h_1^u - h_1^d$ (solid line) plotted against the
longitudinal quark distribution $g_1^u - g_1^d$ (dashed)
computed in the same model.
The contribution of the discrete level to $g_1^u - g_1^d$
is depicted by the dot--dashed line.}
\end{figure}

\newpage
\begin{figure}
 \vspace{-1cm}
\epsfxsize=16cm
\epsfysize=15cm
\centerline{\epsffile{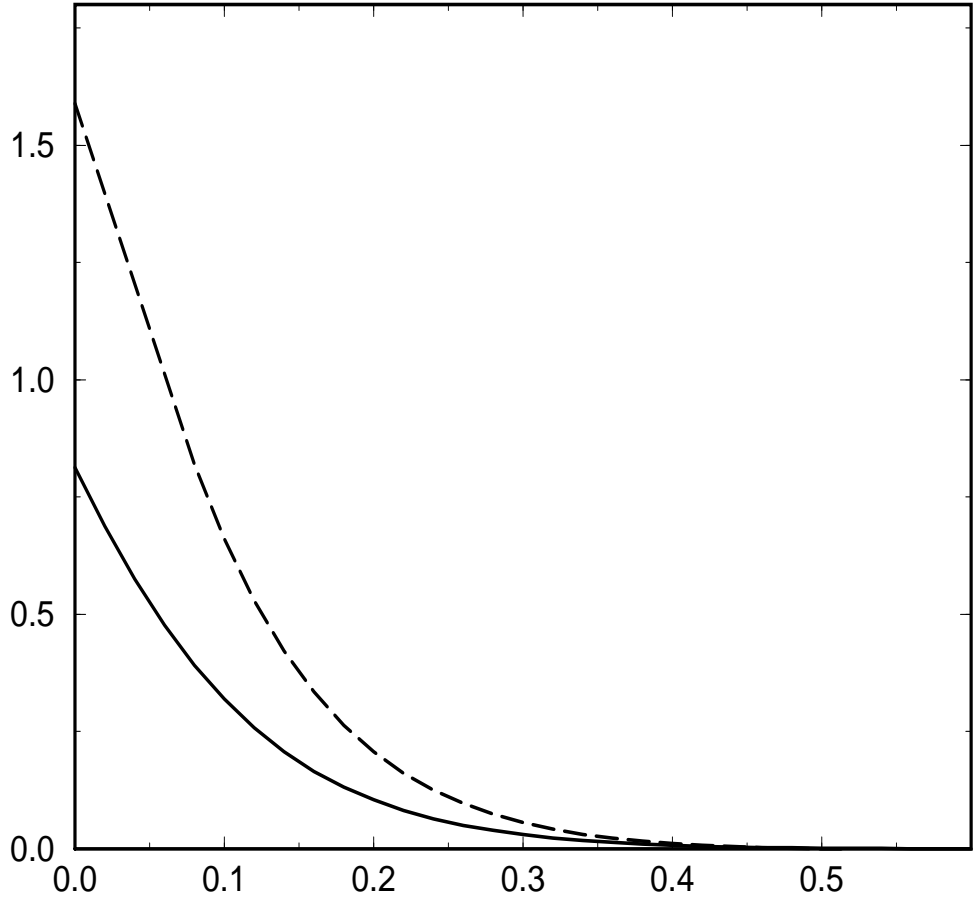}}
\caption[]{
The transverse spin antiquark distribution function
${\bar h}_1^d - {\bar h}_1^u$ (solid line) plotted against the
longitudinal antiquark distribution ${\bar g}_1^u - {\bar g}_1^d$
(dashed) computed in the same model.}
\end{figure}
\end{document}